\documentclass[traditabstract]{aa}

\usepackage{natbib}
\usepackage{amsmath}
\usepackage{graphicx}
\usepackage{wasysym}
\usepackage{txfonts}
\usepackage{xspace}
\usepackage{url}

\newcommand{\fu}{4U\,1909+07\xspace}

\newcommand{\inte}{\textsl{INTEGRAL}\xspace}
\newcommand{\chandra}{\textsl{Chandra}\xspace}

\newcommand{\suzaku}{\textsl{Suzaku}\xspace}
\newcommand{\uhuru}{\textsl{Uhuru}\xspace}
\newcommand{\xte}{\textsl{RXTE}\xspace}

\newcommand{\msun}{\ensuremath{\text{M}_{\odot}}}

\newcommand{\redchi}{\ensuremath{\chi^{2}_\text{red}}}
\newcommand{\cps}{\ensuremath{\mathrm{cts}\,\mathrm{sec}^{-1}}\xspace}

\begin{document}

\title{\fu: a well-hidden pearl}

\titlerunning{\fu: a well-hidden pearl}
\author{
 \mbox{F. F\"urst\inst{1}} \and
 \mbox{I. Kreykenbohm\inst{1}} \and
\mbox{S. Suchy\inst{2}} \and
\mbox{L. Barrag\'an\inst{1}} \and 
\mbox{J. Wilms\inst{1}} \and
\mbox{R. E. Rothschild\inst{2}} \and
\mbox{K. Pottschmidt\inst{3,4}}
}
\authorrunning{F.~F\"urst et al.}

\institute{
 Dr.~Karl Remeis-Sternwarte \& ECAP, Universit\"at Erlangen-N\"urnberg, Sternwartstr.~7, 96049~Bamberg, Germany
\and Center for Astrophysics \& Space Sciences, University of California, San Diego, 9500 Gilman Drive, La Jolla, CA 92093, USA
\and CRESST and NASA Goddard Space Flight Center, Astrophysics Science Division, Code 661, Greenbelt, MD 20771, USA
\and Center for Space Science and Technology, University of Maryland Baltimore County, 1000 Hilltop Circle, Baltimore, MD 21250, USA
}
\date{Received: --- / Accepted: ---}

\abstract{
We present the first detailed spectral and timing analysis of the High Mass X-ray Binary (HMXB) \fu with \inte and \xte. \fu is detected in the ISGRI 20--40\,keV energy band with an average countrate of 2.6\,\cps.
The pulse period of $\sim$604\,sec is not stable, but changing erratically on timescales of years. The pulse profile is strongly energy dependent: it shows a double peaked structure at low energies, the secondary pulse decreases rapidly with increasing energy and above 20\,keV only the primary pulse is visible.
This evolution is consistent between PCA, HEXTE, and ISGRI. 
The phase averaged spectrum can be well described by the sum of a photoabsorbed power law with a cutoff at high energies and a blackbody component. To investigate the pulse profile, we performed phase resolved spectral analysis. We find that the changing spectrum can be best described with a variation of the folding energy. We rule out a correlation between the black body component and the continuum variation and discuss possible accretion geometries.
}

\maketitle

\section{Introduction}
The accreting neutron star X-ray binary \fu was first mentioned in the $3^\text{rd}$ \uhuru catalog as 3U~1912+07 \citep{giacconi74a}. The position and name of the source  were later refined to \fu in the $4^\text{th}$ \uhuru catalog \citep{forman78a}.  Many other X-ray missions, such as \textsl{OSO 7}, \textsl{Ariel 5}, and \textsl{EXOSAT}, among others, also detected a source close to these coordinates. \citet{wen00a} showed that these detections are likely all originating from the same source and refer to it as X1908+075. We will use the name \fu, to honor the original discovery.

Despite the regular detections, the first dedicated paper to discuss \fu was published in 2000 when \citet{wen00a}
analyzed \xte-ASM data and found a stable period of 4.4\,days. This period is interpreted as the orbital period of a binary system. Due to the high photoabsorption, however, no optical counterpart could be identified.  In \xte-PCA data, \citet{levine04a} found a second, shorter period of $\sim$605\,s in the X-ray flux, explained as the pulse period of a slowly rotating neutron star. Using Doppler delay curves, they could also refine the binary orbit parameters and estimated the mass of the companion star to be $M_\star = 9$--$31$\,M$_\odot$ and the radius of the companion  star to be $R_\star \leq 22\,R_\odot$, adopting a canonical mass of 1.4\,M$_\odot$ for the neutron star. One year later, \citet{morel05a} detected an OB star in the near infrared at the location of the X-ray source, thus confirming that the system is a High Mass X-ray Binary (HMXB). The distance of the system was estimated to be 7\,kpc \citep{morel05a}. Prior to this discovery, \citet{levine04a} argued that the companion star could be a Wolf-Rayet star, which would make this system a possible progenitor system to a neutron star-black hole binary. With the identification of the companion, however, this intriguing possibility can be ruled out.
The X-ray luminosity $L$ of \fu is around $2.8\times 10^{36}$\,erg\,s$^{-1}$ for 4.5--200\,keV.
Although the system shows no eclipse, the X-ray flux is still strongly orbital phase dependent \citep{levine04a}. \citet{levine04a} also analyzed orbital phase resolved PCA spectra and found that around orbital phase $\phi_\mathrm{orb} = 1$ the photoabsorption increases by a factor of 2 or more to $N_\mathrm{H} \geq 30\times10^{22}\,\mathrm{cm}^{-2}$, explaining the decreased ASM flux. This increase in absorption can be very well described by a spherical wind model and an inclination of $54^\circ\leq i \leq 70^\circ$, depending on the parameters of the wind model.

In archival \chandra data, \citet{torrejon10a} found that the soft energy spectrum of \fu shows evidence for a Compton shoulder on the soft energy edge of the iron K$\alpha$ line. This feature has so far only been seen in GX~301$-$2 but in no other HMXB with \chandra. \citet{torrejon10a} showed that the Compton shoulder is consistent with the X-ray source being embedded in a Compton thick medium, which is also responsible for the observed photo absorption. Additionally these authors have shown that \fu is an exception from the correlation between $N_\mathrm{H}$ and the equivalent width of the iron line found in most other HXMB \citep{inoue85a}. The origin of this exception is still a mystery and can only be investigated with  more high-resolution spectra.

Although there was renewed interest in \fu in recent years, many basic parameters are still unknown. For example, despite the detection of the pulse period, its evolution with time was not yet studied, even though it can give insight on the prevailing accretion mechanism \citep{bildsten97a}. Additionally, no detailed analysis of the high energy spectrum has been carried out so far. \xte data provide enough statistics to even perform pulse phase resolved spectroscopy, allowing us to obtain a better understanding of the accretion region and mechanism together with the energy dependence of the pulse profile. In this article we are aiming at improving our knowledge in these points, using data from \inte and \xte.
In Sect.~\ref{sec:data} we present the data and reduction methods. The pulse period evolution and the pulse profiles at different energies are analyzed in Sect.~\ref{sec:timing}. In Sect.~\ref{sec:spectra} we perform phase averaged and phase resolved spectroscopy. We summarize and discuss our results in Sect.~\ref{sec:disc}.

\section{Observations and data reduction}
\label{sec:data}
For our study of the \fu system we used all available public data from the X-ray missions INTERnational Gamma-Ray Astrophysics Laboratory \citep[\inte,][]{winkler03a} and Rossi X-Ray Timing Explorer \citep[\xte,][]{bradt93a}. One of the main detectors of \inte is ISGRI, a coded mask instrument sensitive in the 15\,keV -- 1\,MeV energy range and part of the  Imager on Board the Integral Satellite \citep[IBIS;][]{ubertini03a,lebrun03a}. 
IBIS/ISGRI was the first instrument to produce high resolution images of the X-ray sky above 20\,keV. 
Even though no pointed observations on \fu were performed with \inte, thanks to IBIS/ISGRI's large field of view of almost $30^\circ\times30^\circ$, there are more than 8.5\,Msec of off-axis data available. A huge part of these data were obtained in the core program on the Galactic Center performed in the early Announcements of Opportunity \citep[AOs;][]{winkler01a}. Data from later AOs were mainly taken from the extended monitoring campaign of the microquasars GRS~1915$+$105 \citep{rodriguez08a} and SS~433 \citep{krivosheyev09a}. See Table~\ref{tab:obslog} for a detailed list of the data used.

\begin{table}
\caption{Observation log of the \xte and \inte data used. The dates given are the start dates of the respective observation. }
 \begin{tabular}{lcr}
\hline\hline
 Obs-ID / ScW & Start date & Exposure \\\hline
\xte & &\\
     70083-01-01-00 -- 70083-01-28-00 & 2002-12-23 & 105\,ksec \\
  70083-02-01-00 -- 70083-02-19-00 & 2003-01-30 & 75\,ksec \\\hline
\inte & &    \\
004800030010 -- 007000670010 & 2003-03-06 &   1352\,ksec \\
  017200250010 -- 020200020010 & 2004-03-11 & 1192\,ksec \\
  022600370010 -- 025900180010 & 2004-08-20 & 966\,ksec \\
   029100170010 -- 032300140010 & 2005-03-02  & 340\,ksec \\
 036100020010 -- 038300520010 & 2005-09-27 & 880\,ksec \\
047600030010 -- 050600200010 & 2006-09-06 & 397\,ksec \\
   053700090010 -- 055600710010 & 2007-03-07 & 608\,ksec \\
060100090010 -- 062300080010 &  2007-09-15 & 967\,ksec \\
072200090010 -- 074300080010 &  2008-09-11 & 1072\,ksec \\
078200500010  -- 080400080010 &  2009-03-10 & 690\,ksec 
\end{tabular} 

\label{tab:obslog}
\end{table}

\begin{figure}
 \centering
- \includegraphics[width=0.95\columnwidth]{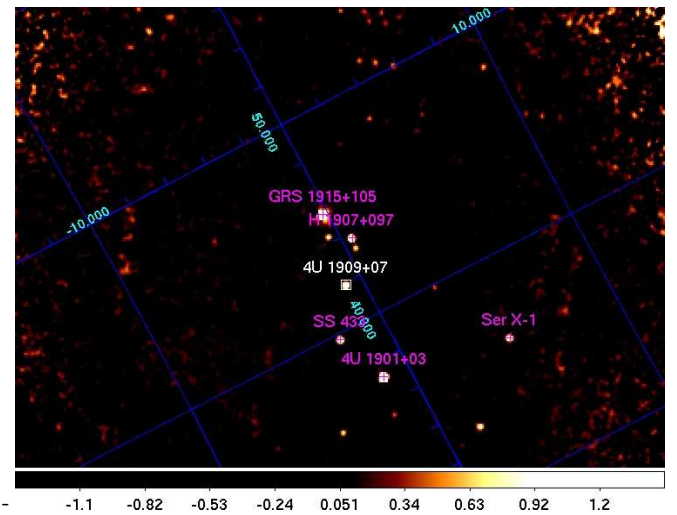}
 \caption{ISGRI image of the X-ray sky in the 20--40\,keV energy band, averaged over 353 ScWs from Revs. 51--68. Coordinates are in galactic longitude and latitude. To avoid confusion only the 6 brightest of overall 12 detected sources are labeled. In this image \fu is the fourth brightest source in the field of view after GRS~1915+105, the transient 4U~1901+03, and H~1907+097.}
 \label{fig:mosaic}
\end{figure}

An IBIS/ISGRI mosaic showing the hard X-ray sky in spring 2003 for the region around \fu is shown in Fig. \ref{fig:mosaic}. 
Twelve sources with a detection significance $> 6\sigma $ were found in that mosaic, with \fu being the fourth brightest source with $\sim$2.6\,\cps.  The data were extracted using the standard pipeline of the Off-line Scientific Analysis software (OSA) 7.0 for spectra and images and using \texttt{ii\_light} to obtain lightcurves with a higher temporal resolution. We extracted lightcurves with a resolution of 20\,sec as a trade-off between good signal-to-noise ratio and good time resolution to measure the pulse period.
We shifted the time to the barycenter of the solar system and corrected for the orbital motion of the neutron star, using the ephemeris given by \citet{levine04a}.

For phase resolved spectroscopy we used \xte observations carried out
in early 2003 which were previously analyzed by \citet{levine04a}.
These data had a total exposure of 196\,ksec. The only other \xte data
are from a shorter observation in 2000, which are not suitable for our
study. Both major instruments of \xte were fully operational at the
time of the observations in 2003, the Proportional Counter Array
\citep[PCA;][]{jahoda06a}, sensitive in the 2--60\,keV energy band and
the High Energy X-Ray Timing Experiment
\citep[HEXTE;][]{rothschild98a}, sensitive between 15--250\,keV. Both
instruments have a large effective area and a high time resolution,
but no imaging capabilities. With a field of view of $\sim$$1^\circ$
no other sources are visible to \xte's instrument while pointing at
\fu. \xte data were reduced using the standard HEASARC software
(v.~6.6), obtaining lightcurves of the PCA with 1\,sec resolution and
pulse phase resolved spectra for PCA and HEXTE in the 4.5--150\,keV
energy range.
The lightcurves were shifted to the barycenter of the solar system and corrected for the neutron star's orbital motion in the same way as the \inte data.

\section{Timing Analysis}
\label{sec:timing}

\subsection{Pulse period evolution}
Monitoring the evolution of the pulse period allows us to determine the type of accretion, i.e., if direct wind accretion prevails or if a stable accretion disk is formed. For \fu the history of pulse period measurements starts only in 2000 with its detection in \xte data \citep{levine04a} and is continued 2 years later by another measurement of \xte. At both times a period of $P \approx 604.7$\,s was measured \citep{levine04a}.
The archival \inte data  provide an optimal basis to perform many more measurements of the pulse period between 2003 and 2008 and to follow its evolution more closely. To do so, we split the IBIS/ISGRI lightcurve into segments between 300\,ksec  and 800\,ksec length, providing a good balance between accurate pulse period determination and temporal resolution, and used the epoch folding technique \citep{leahy87a} with 24 phase bins to determine the pulse period for each of these segments individually. 

\begin{figure}
 \centering
 \includegraphics[width=0.95\columnwidth]{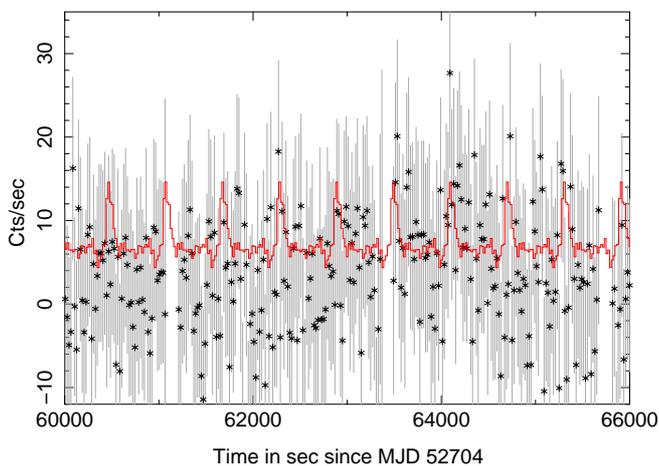}
 \caption{Small part of the lightcurve in the 20--40\,keV energy band as measured with IBIS/ISGRI. Data are shown as black stars with gray bars indicating the uncertainties. Time resolution is 20\,sec. The line shows the folded pulse profile over the whole data set. }
 \label{fig:lc2pprof}
\end{figure}

The counting statistics in ISGRI smear out the single pulses in the lightcurve and make them invisible to the naked eye, see, e.g., Fig. \ref{fig:lc2pprof}.  Only after folding $\approx$500--1000 pulse periods, a stable pulse profile emerges which allows for a reliable determination of the pulse period. This profile is remarkably stable over our data set and not time dependent. 

\begin{figure}
 \centering
 \includegraphics[width=0.95\columnwidth]{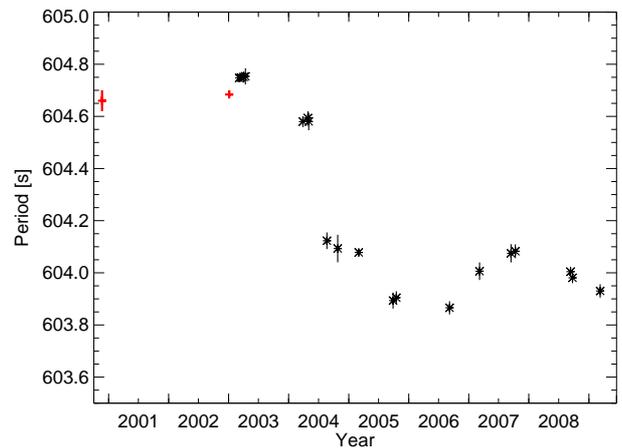}
 \caption{Evolution of the pulse period from 2001 until early 2009. The historic \xte data points of \citet{levine04a} are shown as red crosses, the new measurements obtained with \inte are shown as stars. }
 \label{fig:per_evolu}
\end{figure}

The pulse period evolution between 2001 and 2007 is shown in Fig.~\ref{fig:per_evolu}.
The uncertainties of the pulse period were estimated by fitting a Gaussian distribution to the $\chi^2$ distribution of the epoch folding results. The width of the best-fit Gaussian is a good estimate for the uncertainty of the measurement.

For the pulse period determination we assumed the pulse period to be constant during each individual segment. This assumption is justified, as the  average $\dot{P}$ in our data between two consecutive data points is only $\dot P \approx 1.3\times10^{-8}\,\mathrm{sec\,sec}^{-1}$, very close to the value given by \cite{levine04a}. With the individual lightcurves being on average shorter than 50\,days, the change between start and end of a lightcurve would be only $\Delta P \approx 0.05$\,sec. This change is on the order of the uncertainty we obtained from the Gaussian fits. Even when assuming a larger value of $\dot{P} = 1\,\mathrm{sec\,yr}^{-1} = 3.17\,\mathrm{sec\,sec}^{-1}$ the obtained pulse profiles did not change significantly in shape, making a distinction impossible.

\begin{figure}
 \centering
 \includegraphics[width=0.95\columnwidth]{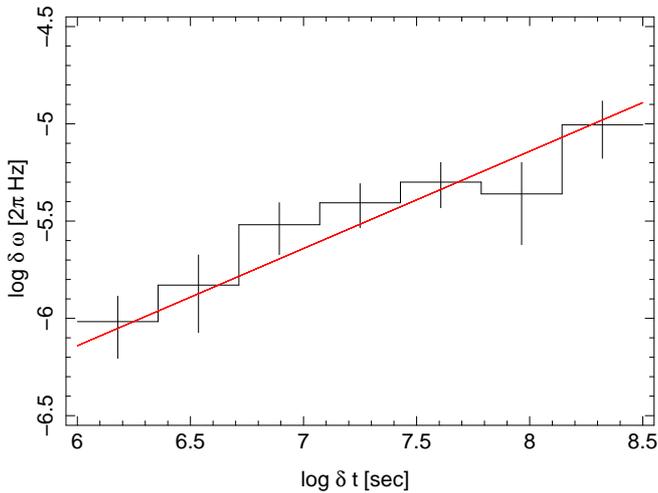}
 \caption{Plot of the pulse period evolution in $\log \delta \omega - \log \delta t$ space, as proposed by \citet{dekool93a}. An ideal random walk process would show a slope of 0.5 in this plot. The data follow this slope closely, as shown by the red straight line with a slope of 0.5.}
 \label{fig:logom_logt}
\end{figure}

The crosses in Fig.~\ref{fig:per_evolu} are \xte measurements taken from \citet{levine04a} and indicate an almost unchanged pulse period between January 2001 and December 2002. Starting with the \inte data, shown as stars in Fig.~\ref{fig:per_evolu}, a spin-up trend between March 2003 and September 2005 is visible from $P\approx604.7$\,sec to to $P\approx 603.9$\,sec. This trend is broken after September 2005 and the pulse period varies around $P = 604.0$\,sec.  These changing trends make the overall behavior consistent with a random walk. To confirm this observation, we used the algorithm proposed by \citet{dekool93a}. It evaluates the relative change in pulse period over different time intervals $\delta t$ between single measurements. To simplify the calculations the pulse period is converted to angular velocity $\omega = {2 \pi}/{P}$ and its change is expressed as the average value of absolute differences:
\[\delta \omega (\delta t)= \left<\left|\omega(t+\delta t) - \omega(t)\right|\right>\]
Each pair ($\omega(t+\delta t),\omega(t)$) is additionally weighted in order to account for the data being grouped in clusters, see \citet{dekool93a} for a detailed description. The results of this analysis are shown in Fig.~\ref{fig:logom_logt}.

If the data are compatible with a random walk, they will follow a straight line with a slope of  0.5 in $\log \delta \omega - \log \delta t$ space. To test if the data are consistent with such a straight line,  we fitted a function of the form 
\[\log(\delta \omega(\delta t)) = A + 0.5\cdot \log(\delta t)\]
where the free parameter $A$ describes the noise level of the data.
The best fit, superimposed in Fig.~\ref{fig:logom_logt}, has a Pearson's correlation coefficient of $r \approx 0.95$ and clearly shows that the pulse period evolution is consistent with a random walk. It gives a noise level of $A = -9.1$, fully in agreement with the expected value for a source with a luminosity of $L \approx  2.8\times 10^{36}$\,erg\,s$^{-1}$, assuming wind accretion with a turbulent accretion wake \citep[and references therein]{dekool93a}.

The uncertainties of the $\delta \omega$-values are originating not only in the determination of the pulse period but also in the uneven and coarse sampling of the pulse period evolution. To estimate the latter effect, we chose a Monte Carlo approach and simulated many mock-up pulse period evolutions which followed a perfect random walk and sampled these with the same rate as the real data \citep[see][]{dekool93a}. The standard deviation in each $\delta t$ bin from these simulations gives an estimate of the uncertainty in each $\delta t$ bin of the data and is roughly of the same order of magnitude as the uncertainties from the pulse period determination.

\subsection{Pulse profiles}
\begin{figure}
 \centering
 \includegraphics[width=0.95\columnwidth]{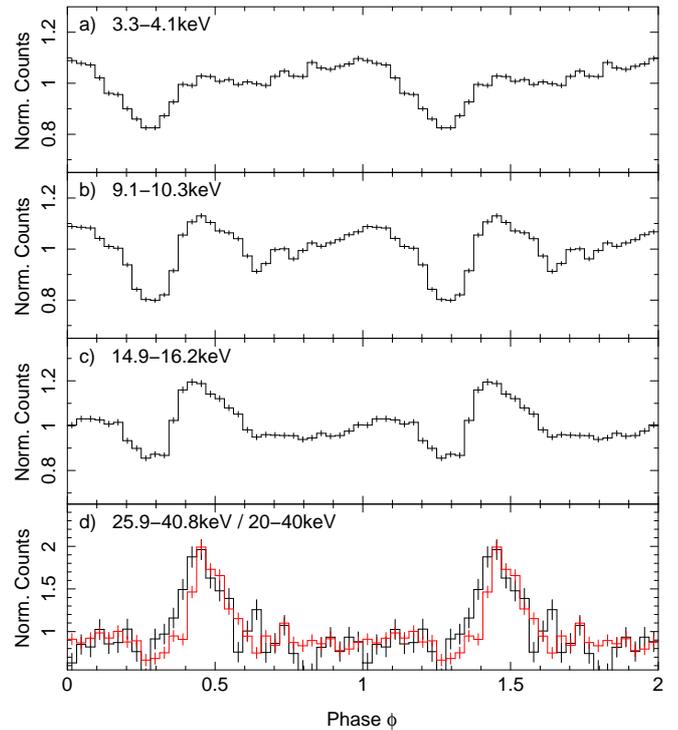}
 \caption{Energy resolved pulse profiles with \xte PCA (a-c) and \xte HEXTE (d). The profiles are shown twice for clarity. In panel d) the \inte pulse profile in the 20--40\,keV energy range is shown in gray for comparison. Note that the \inte profile is shifted by hand to match the peak in the \xte profile, but that the profiles are not phase aligned.}
 \label{fig:pp_energy}
\end{figure}

\citet{levine04a} showed in their Fig.~2 that the average pulse profile between 3.7 -- 17\,keV shows two distinct peaks, with the second one being slightly broader than the first one and exhibiting a complex shape with two subpeaks. 
With higher energy resolution a remarkable and strong energy dependence becomes evident, clearly seen in pulse profiles extracted from \xte/PCA data with a sampling period of $P = 604.685$\,s and 32 phase bins. Thanks to the large effective area of the instrument, very good pulse profiles with a high signal-to-noise ratio could be extracted in 30 narrow energy bands. Three examples are shown in the upper panels of Fig.~\ref{fig:pp_energy}. 
At very low energies below 5\,keV a broad plateau is visible, with increasing intensity at later pulse phases (Fig.~\ref{fig:pp_energy}a). With increasing energy this plateau separates clearly into two peaks, while the second peak is getting dimmer compared to the first one  (Fig.~\ref{fig:pp_energy}b). The relative power of the secondary peak keeps declining up to $\sim$$20$\,keV (Fig.~\ref{fig:pp_energy}c) and it is not visible anymore above 20\,keV (Fig.~\ref{fig:pp_energy}d). Despite the strong evolution with energy of the pulse profile, the deep minimum around phase $\phi = 0.3$ is not energy dependent. The IBIS/ISGRI pulse profile, which was extracted from the average data of the first 100 days of measurement in 2003, is shown in Fig.~\ref{fig:pp_energy}d for comparison. We used the same epoch as for the PCA analysis, but a period of $P = 604.747$\,s, as determined in our analysis of the data. The pulse profile at energies above 20\,keV is consistent between IBIS/ISGRI and \xte/HEXTE.

\begin{figure}
 \centering
 \includegraphics[width=0.95\columnwidth]{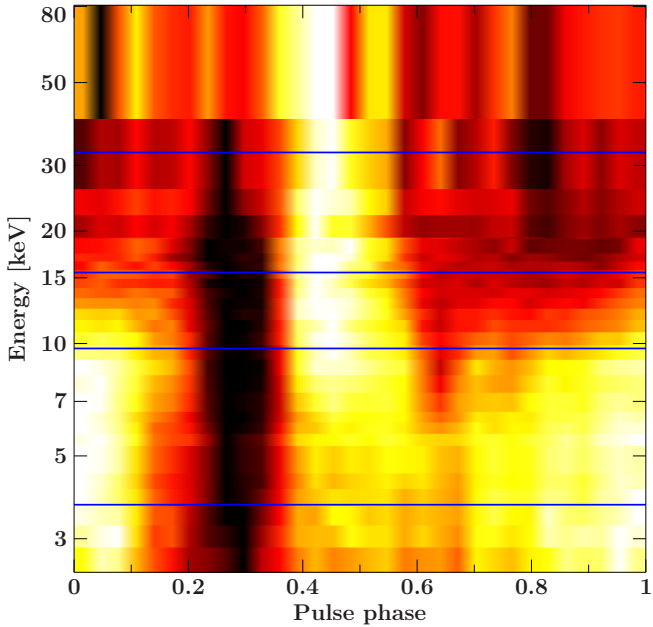}
 \caption{Color coded pulse profile map of PCA and HEXTE data. The color indicates the mean countrate in each bin, ranging from black (lowest) to white (highest). The two highest energy bins are from HEXTE data. The four horizontal blue lines indicate the energies for which the pulse profiles are shown in Fig. \ref{fig:pp_energy}. The map shows one pulse only.}
 \label{fig:pp_landsc}
\end{figure}

To make the shape transition of the pulse profile at low and medium energies better visible, we plotted them in a color coded map (see Fig. \ref{fig:pp_landsc}). The $x$-axis shows the pulse phase, the logarithmically scaled $y$-axis displays the energy in keV and the mean count rate is color-coded. The extensions of the pixels in $y$-direction represent the energy band of the given pulse profile. A smooth transition from the broad, two peaked structure to the single peaked structure is clearly seen. The minimum around phase $\phi= 0.3$ in black stays very stable.

\section{Spectral analysis}
\label{sec:spectra}
\subsection{Phase averaged spectrum}
The pulse phase averaged spectrum of \fu shows a shape similar to many other neutron star sources, consisting of a powerlaw continuum attenuated by photoabsorption at low energies and a turnover at high energies. For this kind of continuum various phenomenological models exist, all describing a similar shape. Among the most widely used models are the cutoff-powerlaw \texttt{cutoffpl}, with a smooth turnover at higher energies (determined by the folding energy $E_\mathrm{fold}$), the \texttt{highecut} model, where the turnover begins suddenly at a given energy $E_\mathrm{cut}$ \citep[and references therein]{white83a} and the Comptonization model \texttt{compTT} \citep{hua95a} with more free parameters to describe the temperature and optical thickness of the Comptonizing plasma. \citet{levine04a} modeled the spectra using a cutoff powerlaw and a bremsstrahlung model, which incorporates only the temperature to describe the spectral shape. They found that both models fit the data equally well.
We applied all four models, \texttt{cutoffpl}, \texttt{highecut}, \texttt{compTT}, and \texttt{bremss} to the \xte and \inte data simultaneously, using ISIS 1.6.0-3 and discarding all observations between orbital phase $0.88 < \phi_\text{orb} < 0.12$, where the $N_\text{H}$ is dramatically increased \citep{levine04a}. The photoabsorption at low energies was modeled using a revised version of the \texttt{tbabs} model\footnote{see \url{http://pulsar.sternwarte.uni-erlangen.de/wilms/research/tbabs/}} with abundances by \citet{wilms00a} and cross-sections by \citet{verner96a}. A strong iron line close to the energy of neutral iron of 6.4\,keV is also presented in the spectrum \citep{levine04a}, and was modeled for all continuum models using a Gaussian emission line.

\begin{figure}
 \centering
 \includegraphics[width=0.95\columnwidth]{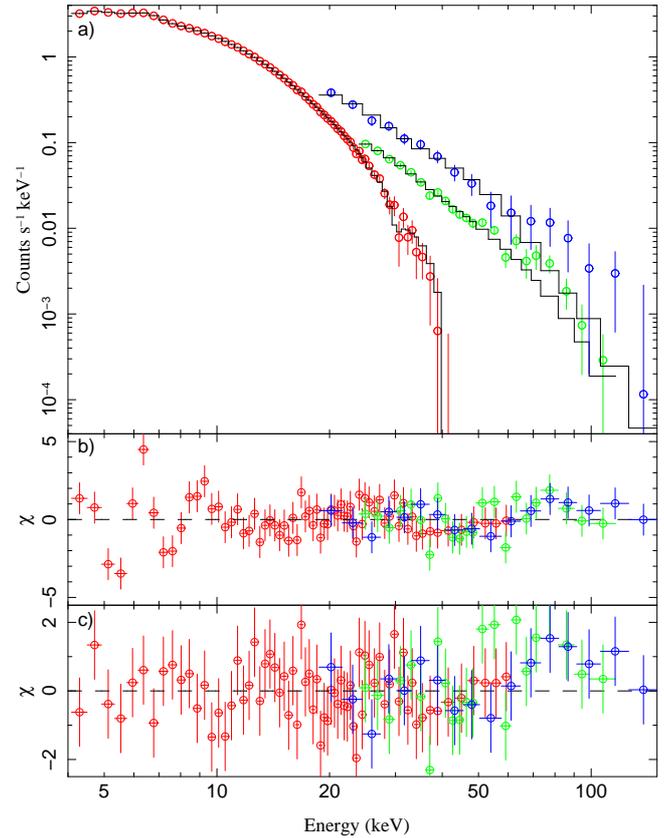}
 \caption{\textit{a)} Combined \xte PCA (red) and HEXTE (green) and \inte ISGRI (blue) spectra. The best fit \texttt{cutoffpl}-model with a blackbody component is shown in black. \textit{b)} Residuals of the best fit model without a black body. \textit{c)} Residuals of the best fit model with a black body.}
 \label{fig:spec_low}
\end{figure}

All continuum models described the data very well, except the bremsstrahlung model, which only gave a best fit reduced $\chi^2$ value of $\redchi = 1.75$. We ascribe this to the fact that the hard X-ray spectrum of \fu can not be modeled accurately with the \texttt{bremss} model, noticeable only thanks to the improved statistics with the additional IBIS/ISGRI data. These data were not available to \citet{levine04a}. The best fit parameters and \redchi values of the other models are presented in Table~\ref{tab:avgpara} and Fig.~\ref{fig:spec_low}a shows the \xte and IBIS/ISGRI spectra together with the best fit \texttt{cutoffpl}. 

The \redchi values are below 1.5 for all models presented in
Table~\ref{tab:avgpara}, indicating a statistically acceptable fit. A
look at the residuals in Fig.~\ref{fig:spec_low}b, however, shows that
the models do not describe the data satisfactorily below 10\,keV.
These residuals can be eliminated by adding
a blackbody component (Fig.~\ref{fig:spec_low}c and Table.~\ref{tab:avgpara}).
The \texttt{cutoffpl} and the \texttt{highecut} model give similar blackbody temperatures between $kT\approx$1.4--1.7\,keV, whereas in the \texttt{compTT} the temperature increased to $kT = 3.1\pm0.4$\,keV. As the $\chi^2$ of the \texttt{compTT + bbody} model is not as good as the $\chi^2$ of the other two models, a temperature around $kT \approx 1.5$\,keV seems most realistic.

\begin{table*}
\caption{Fit parameters for the phase averaged spectra with different models. No significant width of the iron line could be measured for the \texttt{highecut + bbody} and the \texttt{compTT} model and Fe $\sigma$ was consequently frozen to 0.1. Not shown are the cross calibration factors and the renormalization of the PCA background, which are similar for all models and close to 1.}
\label{tab:avgpara}
\centering
\begin{tabular}{lcccccc}
\hline\hline

Model               & cutoffpl & cutoffpl       & highecut & highecut & compTT & compTT \\
parameter           &          & +bbody         &          & +bbody   &        & +bbody  \\\hline
$N_\text{H}$ [$10^{22}$\,cm$^{-2}$]&$20.6\pm1.5$& $14\pm2$ & $15\pm2$ &  $15\pm2$ &  $10.3^{+1.8}_{-1.9}$ & $9\pm2$ \\
$A_\mathrm{cont}$ [ph$@$1\,keV]   &  $0.084^{+0.012}_{-0.011}$  & $0.029^{+0.009}_{-0.007}$ & $0.057^{+0.009}_{-0.011}$ & $0.036^{+0.011}_{-0.009}$ & $(0.54\pm0.03)\times10^{-2}$ & $(4.8\pm0.4)\times10^{-3}$  \\
$\Gamma$& $1.59\pm0.07$ & $1.15\pm0.12$ & $1.56^{+0.06}_{-0.08}$ & $1.36^{+0.11}_{-0.13}$&  --  & -- \\
$E_\mathrm{fold}$ [keV]  &  $28\pm3$ & $20\pm2$ &  $27\pm3$ & $23\pm3$ & -- & -- \\
$E_\mathrm{cutoff}$ [keV]  &  -- & -- & $7.8\pm0.5$  & $7.5^{+0.7}_{-1.0}$ & -- & -- \\
$T_0$ [keV]    & -- & -- & -- & -- & $1.19\pm0.04$ & $1.15\pm0.06$  \\
$kT$ [keV]  & -- & -- & -- & -- & $9.7^{+0.2}_{-0.5}$ & $10.1\pm0.3$  \\
$\tau$    & -- & -- & -- & -- & $3.01^{+0.13}_{-0.01}$ & $3.00^{+0.11}_{-0.00}$  \\
Fe $\sigma$ [keV] & $0.9^{+0.1}_{-0.2}$ & $0.19^{+0.17}_{-0.19}$ & $0.41\pm0.17$ & $0.1$ & $0.1$ & $0.15^{+0.16}_{-0.15}$  \\
Fe Energy [keV] & $6.76\pm0.15$ & $6.47\pm0.07$ & $6.43^{+0.07}_{-0.09}$ & $6.42^{+0.10}_{-0.05}$ & $6.44^{+0.07}_{-0.06}$ & $6.44\pm0.06$\\
Fe A~$\mathrm{[ph/ s/}\mathrm{cm}^{2}\mathrm{]}$& $(0.74^{+0.16}_{-0.15})\times10^{-3}$ & $(4.3^{+1.2}_{-0.9})\times10^{-4}$ & $(0.6^{+0.2}_{-0.1})\times10^{-3}$ & $(3.9^{+0.8}_{-0.7})\times10^{-4}$ & $(3.4^{+0.5}_{-0.4})\times10^{-4}$ & $(4.5^{+1.0}_{-0.7})\times10^{-4}$  \\
bbody $kT$ [keV] &  -- & $1.65^{+0.11}_{-0.08}$ & -- & $1.4^{+0.2}_{-0.1}$ & --  & $3.1\pm0.4$ \\
bbody norm &  -- & $(0.71^{+0.15}_{-0.14})\times10^{-3}$ & -- & $(5\pm2)\times10^{-4}$ & -- & $(3.7^{+1.4}_{-1.3})\times10^{-4}$ \\
$\chi^2/\text{dof}$ $(\chi^2_\text{red})$ & $147.4/100 $ $(1.47)$     & $94.6/98$ $(0.97)$     & $102.7/99$ $(1.04)$     & $85.4/98$ $(0.87)$     &  $124.5/100$ $(1.25)$   & $98.8/97$ $(1.02)$ \\

\end{tabular} 
\end{table*}

\subsection{Phase resolved spectra}
\begin{figure}
 \centering
 \includegraphics[width=0.91\columnwidth]{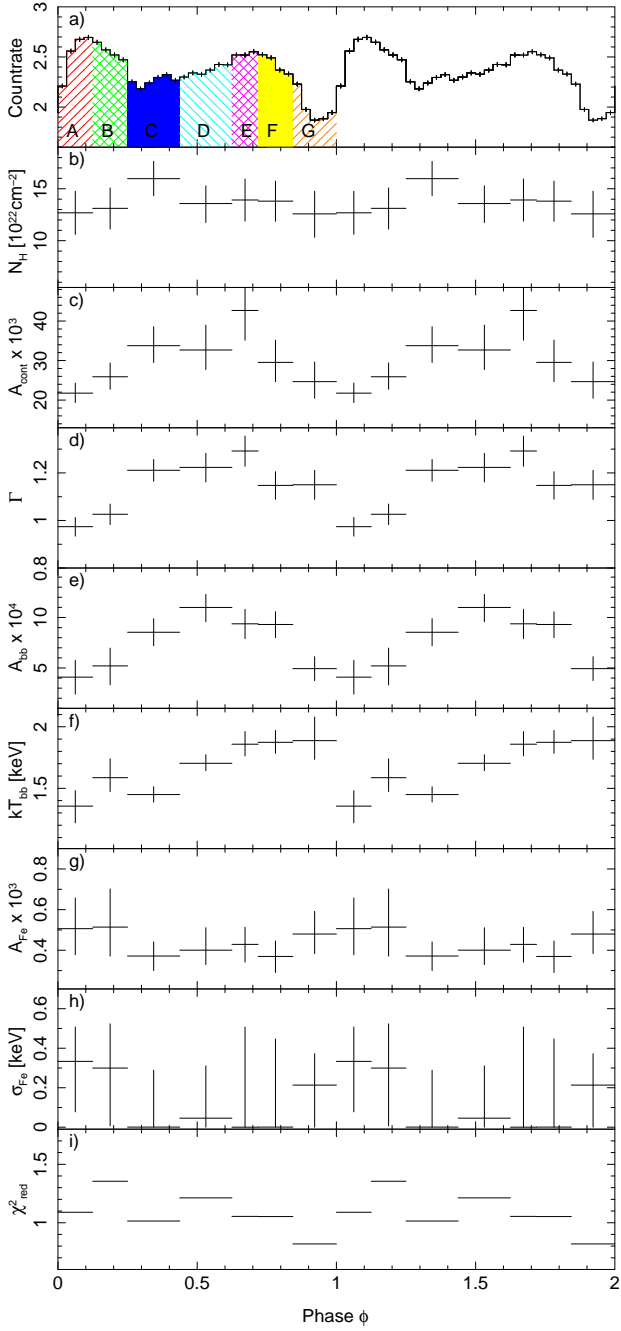}
 \caption{Parameters of the phase-resolved spectra fitted with the \texttt{cutoffpl} model and the folding energy frozen to $E_\mathrm{fold}=20.0$\,keV . \textit{a)}~Pulse profile in the 13.2-14.5\,keV energy range. The different shaded areas indicate the phase bins used to extract the spectra. \textit{b)}~Photo electric absorption column $N_\text{H}$, \textit{c)}~power law normalization, \textit{d)}~power law index $\Gamma$, \textit{e)}~black body normalization, \textit{f)}~black body temperature, \textit{g)}~iron line normalization, \textit{h)}~iron line width, and \textit{i)}~\redchi-value. }
 \label{fig:phasresparas_frozCut}
\end{figure}

Due to the rotation of the neutron star, the accretion columns and hot-spots are seen under a constantly changing viewing angle. Since the physical conditions can not be assumed to be constant throughout the emission regions, the X-ray spectrum can change significantly with pulse phase. These dramatic changes are indicated in the energy dependence of the pulse profile (Fig. \ref{fig:pp_energy} and \ref{fig:pp_landsc}). Pulse phase resolved spectroscopy allows us to disentangle the contributions from the different physical components \citep[see, e.g.,][]{kreykenbohm04b}.
To perform phase resolved spectral analysis, we divided the \xte/PCA and \xte/HEXTE data into 7 phase bins, as indicated in Fig.~\ref{fig:phasresparas_frozCut}a. The IBIS/ISGRI data did not provide good enough statistics for high resolution phase resolved spectroscopy. A \texttt{cutoffpl}, a \texttt{highecut}, and a \texttt{compTT} model, each improved by photoabsorption, an iron line and a black body were fitted to all phase bins.
All three models showed similar behavior as function of pulse phase.
In terms of \redchi, however, the \texttt{highecut} failed in most
phase bins very clearly, even though it gave the best \redchi values
for the phase averaged spectrum. A possible explanation for this
effect is that the phase averaged spectrum is a superposition of the
different spectra at different phases and that this superposition is
best described with the \texttt{highecut} model. To check this
assumption, we simulated a spectrum consisting of a superposition of
the best fit \texttt{cutoffpl} models for the individual phasebins,
each parameter weighted by the respective count rate in the respective
phase bin. The emerging spectrum could be described significantly
better with a \texttt{highecut} model ($\redchi = 0.72$) than with the
\texttt{cutoffpl} model ($\redchi = 0.82$). This result shows that the
superposition of the simpler \texttt{cutoffpl} models introduces new
features in the phase averaged spectrum, which are not source
intrinsic. The \texttt{cutoffpl} model gave the best fits over all
phasebins and only those results are presented here.

Describing the spectra without a blackbody component resulted in good \redchi values only for phase bins A, B, C, and G, but returned  formally unacceptable \redchi values for phase bins D, E, and F, i.e., for spectra during the secondary peak. As discussed in the previous section the phase averaged spectrum can also be fitted relatively well without a black body component, but the inclusion of such a component improves the $\chi^2$ value. This effect can now be explained by viewing the phase averaged spectrum as a superposition of spectra with and without a prominent black body component, and consequently showing only comparatively weak evidence for this component. Phase resolved analysis disentangles this superposition and gives more information on the real physical behavior of the source.

A close analysis of all spectra revealed that the folding energy and the photon index of the \texttt{cutoffpl} model depend strongly on each other and that the statistics of the spectra do not allow us to separate the influence of these two parameters. We consequently carried out two independent fits, one with the folding energy frozen to its value in the phase averaged fit and one with a frozen photon index, also set to the value of the phase averaged fit. Both models gave 
 very good results in terms of \redchi. To further reduce the number of variable parameters, we also froze the energy of the Gaussian iron line to 6.47\,keV, the value of the phase averaged fit. The iron line energy was consistent with that value in all phase bins when allowed to vary. 
 Overall the free parameters of the fit were: the normalization of the \texttt{cutoffpl}, the column of the photo absorption, the normalization and temperature of the black body, the normalization and width of the iron line and either the folding energy or the photon index of the \texttt{cutoffpl}.

\begin{figure}
 \centering
 \includegraphics[width=0.91\columnwidth]{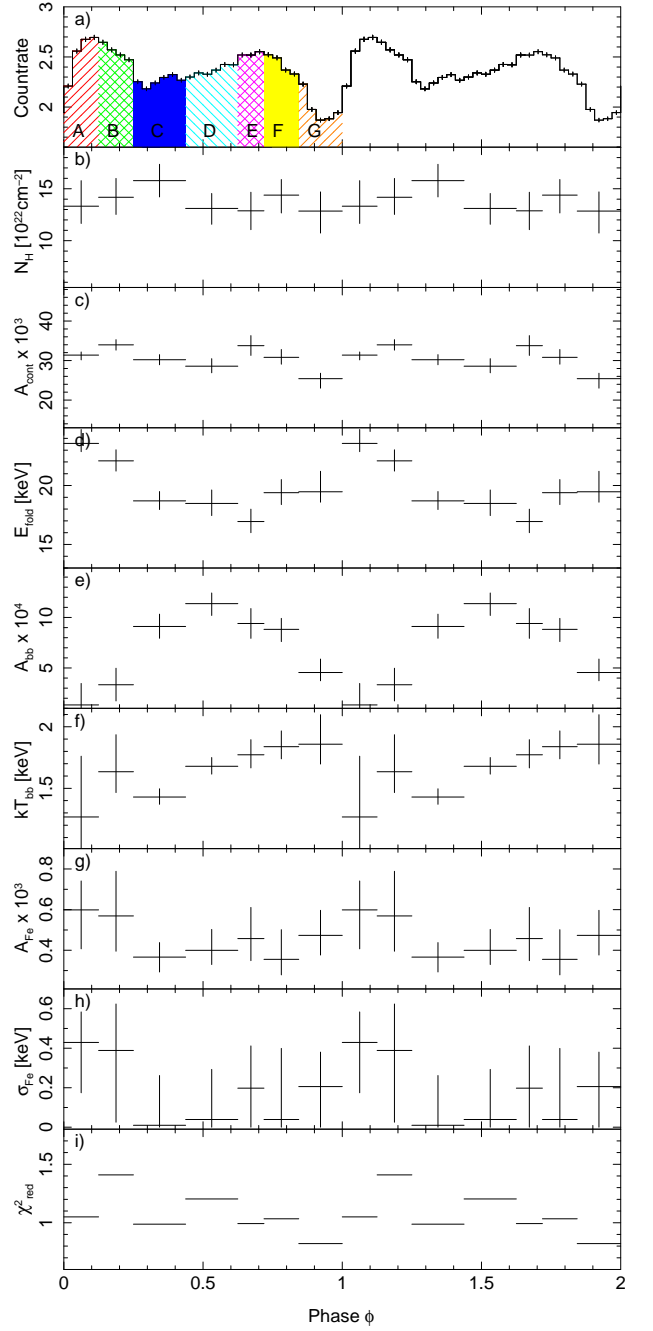}
 \caption{Same as Fig.~\ref{fig:phasresparas_frozCut}, but for the \texttt{cutoffpl} model with the photon index frozen to $\Gamma = 1.15$. \textit{a)}~Pulse profile in the 13.2-14.5\,keV energy range, \textit{b)}~photo electric absorption column $N_\text{H}$, \textit{c)}~power law normalization, \textit{d)}~folding energy, \textit{e)}~black body normalization, \textit{f)}~black body temperature  \textit{g)}~iron line normalization, \textit{h)}~iron line width, and \textit{i)}~\redchi-value.}
 \label{fig:phasresparas_frozPho}
\end{figure}

The evolution of the spectral parameters is shown in Fig. \ref{fig:phasresparas_frozCut} for a frozen folding energy and in Fig.~\ref{fig:phasresparas_frozPho} for a frozen photon index. It is clearly seen that for both models neither the photoabsorption $N_\text{H}$ nor the power of the iron line seem to vary much over the pulse phase. The strong spectral changes indicated in the energy resolved pulse profiles are thus mainly due to a change in the continuum of the \texttt{cutoffpl} model, either due to an increase in the photon index (Fig. \ref{fig:phasresparas_frozCut}d) or due to a shift of the folding energy to higher energies (Fig. \ref{fig:phasresparas_frozPho}d). The folding energy is highest in the primary peak during phase bins A and B, making its spectrum distinctly harder than the rest of the pulse phase. In the secondary pulse the folding energy moves to values as low as $\sim$17\,keV, forcing a strong attenuation of the spectrum above this energy.

The behavior of the black body component is also very interesting in both models. Figures \ref{fig:phasresparas_frozCut} and \ref{fig:phasresparas_frozPho} show the normalization of the black body component and its temperature (panels e and f, respectively). While the temperature varies insignificantly with pulse phase, the normalization, expressing the contribution of the black body component to the overall flux, shows a strong dependency on pulse phase with variation up to a factor of 3.  The black body is strongest in phase bin D, i.e., during the beginning of the secondary pulse, and lowest in phase bin A, i.e., during the rise and maximum of the primary peak. As there is an indication for a broadening of the iron line during the phase bins where the black body is weakest, one might speculate about a compensation of the blackbody component by the Gaussian.  Confidence contours between the two parameter, however, showed no correlation. The indicated broadening of the iron line is more likely due to a change in the iron edge or an iron K$\beta$ line, which can not be resolved with \xte/PCA. Using spectra with higher energy resolution a more in depth study of the iron line behavior with pulse phase will be possible.

Regarding the normalization of the powerlaw, it clearly reflects the pulse profile in the model where the photon index was frozen (see Fig.~\ref{fig:phasresparas_frozPho}c). In the model with the frozen folding energy, on the other hand, the correlation is not so clear. It rather seems that in this case a combination of the photon index and the power law normalization models not only the spectral change, but also the brightness variation. Because of this entanglement, the model with the variable folding energy seems the more realistic description of the spectrum.

\begin{figure}
 \centering
 \includegraphics[width=0.93\columnwidth]{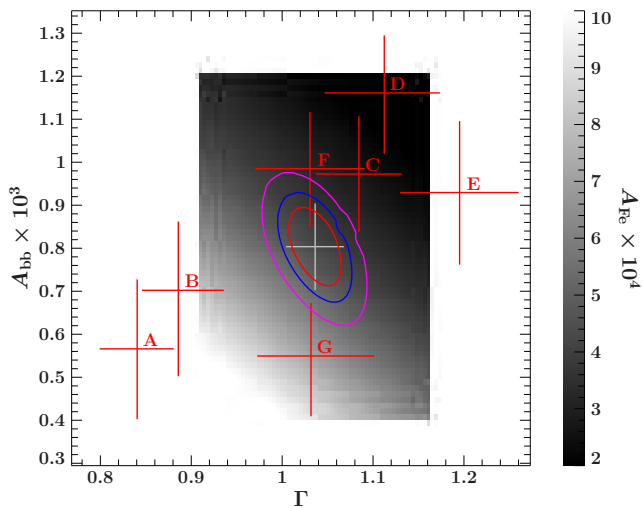}
 \caption{Confidence contour and best fit value for the phase averaged spectrum between the photon index of the powerlaw and the normalization of the blackbody component, when fitted with a frozen folding energy. The contour lines show from the center outwards 68\% (1$\sigma$), 90\% and 99\% confidence level. Gray-scaled is the normalization of the Gaussian iron line for the best-fit model at every given photon index -- blackbody normalization point. The points labeled A through G are the best fit values of the respective phase bins and their uncertainties in the same space and for the same model.}
 \label{fig:confcontgamma2bb}
\end{figure}

To exclude strong dependencies we calculated confidence contours for the phase averaged spectrum for the blackbody normalization versus the powerlaw normalization as well as for the blackbody normalization versus the folding energy, based on the models with frozen folding energy and frozen powerlaw normalization, respectively.
Figure~\ref{fig:confcontgamma2bb} shows the confidence contour for the model with the variable photon index, i.e., the frozen folding energy, together with the best fit value as a bright cross. The contours are indicating only a very weak correlation between the black body normalization and the photon index. In the background the best fit value of the normalization of the Gaussian iron line is shown in gray-scale for every relevant point in the black body normalization -- photon index space. The scale on the right of the plots gives its corresponding values. With increasing blackbody power and spectral softening a weaker iron line is necessary. Within the best fit contours, however, the iron line is not changing dramatically, thus ruling out a substitution of the blackbody with a stronger iron line for formally acceptable \redchi values.

Additionally to the phase averaged contour, Fig.~\ref{fig:confcontgamma2bb} shows the best fit values for all 7 phase bins as dark crosses, labeled with the respective letter. As can be clearly seen, in phase bins A and B the spectrum is hardest, while the black body component is weakest. Phase bins C--F are clustered together in a regime with a slightly softer spectrum and stronger black body than the phase averaged spectrum. Phase bin G, describing the minimum between two consecutive pulses shows only very weak portions of the black body and an intermediate spectral hardness. This distribution makes very clear that both, the photon index and the black body component change significantly over the pulse phase in a way which seems to indicate a correlation between them. This correlation is most likely of physical origin and not introduced due to improper modeling of the spectrum, as the confidence contour of the phase averaged spectrum showed no correlation between these parameters. 

\begin{figure}
 \centering
 \includegraphics[width=0.93\columnwidth]{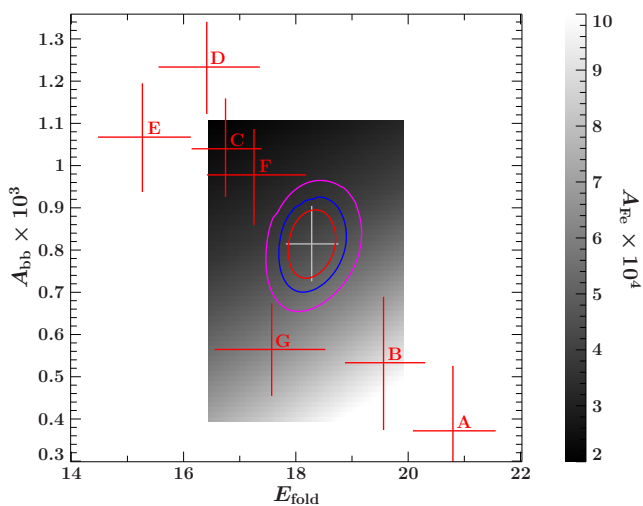}
 \caption{Same as Fig.~\ref{fig:confcontgamma2bb} but for the folding energy of the power law versus the normalization of the black body, while the photo index of the \texttt{cutoffpl} model was frozen.}
 \label{fig:confcontcut2bb}
\end{figure}

Figure~\ref{fig:confcontcut2bb} shows the same plot as Fig.~\ref{fig:confcontgamma2bb} but for the model with a variable folding energy, i.e., a frozen photon index. 
As for the variable photon index, the iron line does not change significantly within the acceptable $\chi^2$ contours and is only a weak function of both parameters. The best fit phase resolved values show a similar distribution as in Fig.~\ref{fig:confcontgamma2bb}.
It is clearly seen that phase bin A shows the hardest spectrum with a negligible black body component, while phase bin B moves a bit to a softer spectrum with a stronger black body. Phase bins C--F are again clustered together and phase bin G shows intermediate spectral hardness with a very weak black body. This distribution of the phase resolved data indicates an anti-correlation between the black body normalization and the folding energy. As in the other model, this correlation is most likely intrinsic to the source as the phase averaged confidence contours rule out a model dependent correlation.

\section{Discussion \& Conclusion}
\label{sec:disc} 
We have presented the first detailed study with \inte and \xte of \fu and have shown that the pulse period evolution shows strong indication for a random walk like behavior. Such a behavior has been seen in other HMXB like {Vela~X-1} and is a strong indicator for a wind accreting source without a persistent accretion disk \citep{ghosh79a}. An accretion disk would provide a more constant transfer of angular momentum and thus a long-term spin-up or spin-down trend, as seen in other sources like 4U~1907+09 \citep{fritz06a}. Accretion disks form to fulfill the conservation of angular momentum of the accreted wind, but in a system with a strongly magnetized neutron star the ionized matter will couple to the magnetic field lines before a stable disk can form. This requires field strengths on the order of $10^{12}$\,G, not unusual for this kind of system. 
In these strong magnetic fields cyclotron resonant scattering features (CRSFs) can form, seen in many other sources, e.g.,  Vela~X-1 \citep{kreykenbohm99a}, 4U~0115+63 \citep{santangelo99a}, or MXB~0656$-$072 \citep{mcbride06a}. We found no evidence for a CRSF in the spectra of \fu. As shown by \citet{schoenherr07a} this does not rule out a strong magnetic field, as CRSFs can be filled by photon spawning and their shapes and depths depend strongly on the geometry of the accretion column and on the viewing angle. Further investigations with high-resolution data will allow a more thorough search for a CRSF in \fu.

A strong magnetic field in \fu is supported by the strongly pulsed flux, as the X-ray flux is believed to be produced in a very small region above the hot-spots on the neutron star surface, where it is penetrated by the magnetic field lines and the matter is accreted onto the neutron star \citep{lamb73a}. In the accretion column above the hot-spots, the density and temperatures are highest and the most X-rays are produced through Comptonization of the thermal photons from the thermal mound on the neutron star surface \citep{becker05a, becker05b}. If the magnetic axis is not aligned with the rotational axis of the neutron stars, the hot spots and the accretion columns move in and out of sight of the observer, resulting in a pulsed X-ray flux.

The exact geometry of the accretion column depends strongly on the way the matter couples to the magnetic field and can not be constrained from our measurements \citep{basko76a, meszaros84c}. Besides the simple filled funnel other possible configurations include a hollow or a partly hollow funnel. With an average luminosity of $\sim$$2.8\times 10^{36}$\,erg\,s$^{-1}$ between 4.5--200\,keV the system can host accretion rates that are large enough for a shock to form in the accretion column, due to the radiation pressure from the material close to the neutron star surface. This shock prevents escape of the Comptonized photons parallel to the magnetic field lines, so that they can only emerge from the sides of the accretion column in a so-called ``fan beam'' \citep{meszaros84c}. If the accretion rate is lower, no shock will form and the radiation can escape along the accretion column in the ``pencil beam''.

\subsection{Classical picture}
In a classical assumption the two distinct peaks in the pulse profile at low energies hint at the fact that accretion happens onto both magnetic poles of the neutron star, but under different physical conditions. These differences can explain the different spectra in the two peaks. The secondary peak shows a distinctly lower folding energy which corresponds roughly to the temperature of the electron gas in the accretion column. A misalignment between the magnetic center and the gravitational center of the neutron star could  lead to different sizes of the accretion column and thus to different densities and temperatures. 

We have also seen that the blackbody component of the spectrum of \fu is phase shifted with respect to the two peaks in the pulse profiles.  It is likely that the blackbody originates from the thermal photons of the thermal mound at the hot spot. Through the rotation of the neutron star, the hot spot is only visible at certain pulse phases. Assuming a hollow funnel accretion geometry, with a large enough accretion rate to form a shock, we can argue that we see the thermal mound only at the phases were we look along the hollow accretion column, while the most hard X-ray flux is only visible when we see the sides of the accretion column, due to the ``fan beam'' geometry. As we averaged every phase bin over a large range of time, we can neither rule out nor investigate a change of the accretion geometry.

\subsection{Strongly relativistic picture}
Caution has to be taken, however, with these simple classical models as the behavior of plasma in magnetic fields on the order of $B\approx 10^{12}$\,G is still not well understood and all models are based on strong simplifications. Additionally gravitational light bending must be taken into account when analyzing pulse profiles and their origin \citep{kraus03a, meszaros85b}. Close to the surface of a canonical neutron star of mass $M=1.4\,\msun$ and radius $R=10$\,km, i.e., a radius only 2.4 times the Schwarzschild radius, light bending leads to a visible surface of the neutron star of 83\%. This huge amount of always visible surface reduces the pulsed flux from the two hot-spots. Assuming a ``fan-beam'' emission characteristic of the accretion column, gravitational light bending can increase the pulsed flux dramatically, especially at phases where one column would be hidden behind the neutron star in the classical picture \citep{riffert88a}. In a very special configuration where the rotational axis takes an angle of $45^\circ$ to both the observer and the magnetic axis, pulse profiles very similar to the high energetic pulse profile of \fu (Fig.~\ref{fig:pp_energy}d) can emerge, in which only one sharp peak is visible \citep[compare Fig.~5 of ][]{riffert88a}. In the simple models of \citet{riffert88a} the emission characteristic of the accretion column is taken to be isotropic and independent of energy, so that a direct translation to \fu is not easily made. Attempts to decompose the pulse profile for other sources were made by \citet{kraus03a} and \citet{sasaki10a}, among others. These decompositions prove difficult due to the many free parameters, like emission pattern, geometry of the accretion column, configuration of the magnetic field, and equation of state which can not be closely constrained. Nonetheless the sharp peaked pulse profile of \fu above 20\,keV is a hint that the hard X-ray emission is produced in a ``fan-beam'' accretion column and that during the rotation of the neutron star one column vanishes behind the star and its emission is bent around and focused due to the strong gravitation.

More detailed models for pulse decomposition will be applied to \fu in the future, together with data from other X-ray satellites that provide a higher spectral resolution, such as \suzaku. With these data a detailed investigation of the iron line complex and the Compton shoulder will also be possible and lead to more insight on the physical state of the absorbing medium.

\acknowledgements
We thank the anonymous referee for her/his useful comments. This work was supported by the Bundesministerium f\"ur Wirtschaft und Technologie through DLR grant 50\,OR\,0808 and via a DAAD fellowship. This work has been partially funded by the European Commission under the 7th Framework Program under contract ITN\,215212. FF thanks the colleagues at UCSD and GSFC for their hospitality. This research has made use of NASA's Astrophysics Data System. This work is based on observations with \inte, an ESA project with instruments and science data centre funded by ESA member states (especially the PI countries: Denmark, France, Germany, Italy, Switzerland, Spain), Czech Republic and Poland, and with the participation of Russia and the USA. For this work we used the ISIS software package provided by MIT. We especially like to thank J. C. Houck and J. E. Davis for their restless work to improve ISIS and S-Lang.

\bibliographystyle{jwaabib}

\begin{thebibliography}{}

\bibitem[\protect\astroncite{{Basko} \& {Sunyaev}}{1976}]{basko76a}
{Basko} M.M., {Sunyaev} R.A.,  1976, MNRAS 175, 395

\bibitem[\protect\astroncite{{Becker} \& {Wolff}}{2005a}]{becker05a}
{Becker} P.A., {Wolff} M.T.,  2005a, ApJ 621, L45

\bibitem[\protect\astroncite{{Becker} \& {Wolff}}{2005b}]{becker05b}
{Becker} P.A., {Wolff} M.T.,  2005b, ApJ 630, 465

\bibitem[\protect\astroncite{{Bildsten} et~al.}{1997}]{bildsten97a}
{Bildsten} L., {Chakrabarty} D., {Chiu} J., et~al., 1997, ApJS 113, 367

\bibitem[\protect\astroncite{Bradt et~al.}{1993}]{bradt93a}
Bradt H.V., Rothschild R.E., Swank J.H.,  1993, A\&AS 97, 355

\bibitem[\protect\astroncite{de~Kool \& Anzer}{1993}]{dekool93a}
de~Kool M., Anzer U.,  1993, MNRAS 262, 726

\bibitem[\protect\astroncite{Forman et~al.}{1978}]{forman78a}
Forman W., Jones C., Cominsky L., et~al., 1978, A\&AS 38, 357

\bibitem[\protect\astroncite{Fritz et~al.}{2006}]{fritz06a}
Fritz S., Kreykenbohm I., Wilms J., et~al., 2006, A\&A 458, 885

\bibitem[\protect\astroncite{Ghosh \& Lamb}{1979}]{ghosh79a}
Ghosh P., Lamb F.K.,  1979, ApJ 234, 296

\bibitem[\protect\astroncite{Giacconi et~al.}{1974}]{giacconi74a}
Giacconi R., Murray S., Gursky H., et~al., 1974, A\&AS 27, 37

\bibitem[\protect\astroncite{Hua \& Titarchuk}{1995}]{hua95a}
Hua X., Titarchuk L.,  1995, ApJ 449, 188

\bibitem[\protect\astroncite{Inoue}{1985}]{inoue85a}
Inoue H.,  1985, Space Sci. Rev. 40, 317

\bibitem[\protect\astroncite{{Jahoda} et~al.}{2006}]{jahoda06a}
{Jahoda} K., {Markwardt} C.B., {Radeva} Y., et~al., 2006, ApJS 163, 401

\bibitem[\protect\astroncite{Kraus et~al.}{2003}]{kraus03a}
Kraus U., Zahn C., Weth C., Ruder H.,  2003, ApJ 590, 424

\bibitem[\protect\astroncite{{Kreykenbohm} et~al.}{1999}]{kreykenbohm99a}
{Kreykenbohm} I., {Kretschmar} P., {Wilms} J., et~al., 1999, A\&A 341, 141

\bibitem[\protect\astroncite{Kreykenbohm et~al.}{2004}]{kreykenbohm04b}
Kreykenbohm I., Wilms J., Coburn W., et~al., 2004, A\&A 427, 975

\bibitem[\protect\astroncite{Krivosheyev et~al.}{2009}]{krivosheyev09a}
Krivosheyev Y.M., Bisnovatyi-Kogan G.S., Cherepashchuk A.M., Postnov K.A.,
  2009,
\newblock In: Choliy V.Y., Ivashchenko G. (eds.) YSC'16 Proceedings of
  Contributed Papers., Kyiv, p.49

\bibitem[\protect\astroncite{{Lamb} et~al.}{1973}]{lamb73a}
{Lamb} F.K., {Pethick} C.J., {Pines} D.,  1973, ApJ 184, 271

\bibitem[\protect\astroncite{Leahy}{1987}]{leahy87a}
Leahy D.A.,  1987, A\&A 180, 275

\bibitem[\protect\astroncite{{Lebrun} et~al.}{2003}]{lebrun03a}
{Lebrun} F., {Leray} J.P., {Lavocat} P., et~al., 2003, A\&A 411, L141

\bibitem[\protect\astroncite{Levine et~al.}{2004}]{levine04a}
Levine A.M., Rappaport S., Remillard R., Savcheva A.,  2004, ApJ 617, 1284

\bibitem[\protect\astroncite{{McBride} et~al.}{2006}]{mcbride06a}
{McBride} V.A., Wilms J., Coe M.J., et~al., 2006, A\&A 451, 267

\bibitem[\protect\astroncite{M{\'e}sz{\'a}ros}{1984}]{meszaros84c}
M{\'e}sz{\'a}ros P.,  1984, Space Sci. Rev. 38, 325

\bibitem[\protect\astroncite{{M{\'e}sz{\'a}ros} \& {Nagel}}{1985}]{meszaros85b}
{M{\'e}sz{\'a}ros} P., {Nagel} W.,  1985, ApJ 299, 138

\bibitem[\protect\astroncite{Morel \& Grosdidier}{2005}]{morel05a}
Morel T., Grosdidier Y.,  2005, MNRAS 356, 665

\bibitem[\protect\astroncite{Riffert \& M{\'e}sz{\'a}ros}{1988}]{riffert88a}
Riffert H., M{\'e}sz{\'a}ros P.,  1988, ApJ 325, 207

\bibitem[\protect\astroncite{Rodriguez et~al.}{2008}]{rodriguez08a}
Rodriguez J., Hannikainen D.C., Shaw S.E., et~al., 2008, ApJ 675, 1436

\bibitem[\protect\astroncite{{Rothschild} et~al.}{1998}]{rothschild98a}
{Rothschild} R.E., {Blanco} P.R., {Gruber} D.E., et~al., 1998, ApJ 496, 538

\bibitem[\protect\astroncite{Santangelo et~al.}{1999}]{santangelo99a}
Santangelo A., Segreto A., Giarrusso S., et~al., 1999, ApJ 523, L85

\bibitem[\protect\astroncite{Sasaki et~al.}{2010}]{sasaki10a}
Sasaki M., Klochkov D., Kraus U., et~al., 2010, A\&A 517, 8

\bibitem[\protect\astroncite{{Sch{\"o}nherr} et~al.}{2007}]{schoenherr07a}
{Sch{\"o}nherr} G., {Wilms} J., {Kretschmar} P., et~al., 2007, A\&A 472, 353

\bibitem[\protect\astroncite{Torrej{\'o}n et~al.}{2010}]{torrejon10a}
Torrej{\'o}n J.M., Schulz N.S., Nowak M.A., Kallman T.R.,  2010, ApJ 715, 947

\bibitem[\protect\astroncite{{Ubertini} et~al.}{2003}]{ubertini03a}
{Ubertini} P., {Lebrun} F., {Di Cocco} G., et~al., 2003, A\&A 411, L131

\bibitem[\protect\astroncite{Verner et~al.}{1996}]{verner96a}
Verner D.A., Ferland G.J., Korista K.T., Yakovlev D.G.,  1996, ApJ 465, 487

\bibitem[\protect\astroncite{Wen et~al.}{2000}]{wen00a}
Wen L., Remillard R.A., Bradt H.V.,  2000, ApJ 532, 1119

\bibitem[\protect\astroncite{White et~al.}{1983}]{white83a}
White N.E., Swank J.H., Holt S.S.,  1983, ApJ 270, 711

\bibitem[\protect\astroncite{{Wilms} et~al.}{2000}]{wilms00a}
{Wilms} J., {Allen} A., {McCray} R.,  2000, ApJ 542, 914

\bibitem[\protect\astroncite{Winkler}{2001}]{winkler01a}
Winkler C.,  2001,
\newblock In: Battrick B. (ed.) Proc. Fourth INTEGRAL Workshop. ESA SP-459, ESA
  Publications Division, Noordwijk, p.471

\bibitem[\protect\astroncite{{Winkler} et~al.}{2003}]{winkler03a}
{Winkler} C., {Courvoisier} T.J.L., {Di Cocco} G., et~al., 2003, A\&A 411, L1

\end{thebibliography}

\end{document}